\documentclass[aps, prc, nofootinbib,
superscriptaddress, twocolumn,10pt,nopreprintnumbers,noeprint]{revtex4-2}
\usepackage[utf8]{inputenc}
\usepackage[dvipsnames]{xcolor}
\usepackage{amsthm, amsfonts}
\usepackage{slashed}
\usepackage{physics}
\usepackage{xspace}
\usepackage{makecell}\usepackage{comment}
\usepackage{orcidlink}
\usepackage{hyperref}

\hypersetup{
    colorlinks,
    linkcolor={red!50!black},
    citecolor={blue!50!black},
    urlcolor={blue!80!black}
}

\usepackage{braket}
\usepackage{multirow}
\usepackage{bbm}
\usepackage{bm}  %
\usepackage{placeins}
\usepackage{cancel}

\usepackage{glossaries}
\setacronymstyle{long-short}
\glsdisablehyper
\newacronym{ChiEFT}{\ensuremath{\chi}\text{EFT}}{chiral effective field theory}
\newacronym{lo}{LO}{leading order}
\newacronym{nlo}{\ensuremath{\text{NLO}}}{next-to-leading order}
\newacronym{n2lo}{\ensuremath{\text{NN}\text{LO}}}{next-to-next-to-leading order}
\newacronym{n3lo}{\ensuremath{\text{NNN}\text{LO}}}{next-to-next-to-next-to-leading order}
\newacronym{n4lo}{\ensuremath{\text{NNNN}\text{LO}}}{next-to-next-to-next-to-next-to-leading order}

\newcommand{\nopieft}{$\cancel{\pi}$EFT\xspace}
\newcommand{\pieft}{$\chi$EFT\xspace}
\newcommand{\q}{\textbf{\emph{q}}}
\newcommand{\NN}{\ensuremath{N\!N}\xspace}

\DeclareMathOperator*{\SumInt}{%
\mathchoice%
  {\ooalign{$\displaystyle\sum$\cr\hidewidth$\displaystyle\int$\hidewidth\cr}}
  {\ooalign{\raisebox{.14\height}{\scalebox{.7}{$\textstyle\sum$}}\cr\hidewidth$\textstyle\int$\hidewidth\cr}}   
  {\ooalign{\raisebox{.2\height}{\scalebox{.6}{$\scriptstyle\sum$}}\cr$\scriptstyle\int$\cr}}
  {\ooalign{\raisebox{.2\height}{\scalebox{.6}{$\scriptstyle\sum$}}\cr$\scriptstyle\int$\cr}}  
}

\begin{document}

\title{Improved nuclear-structure corrections to the hyperfine splitting of electronic and muonic deuterium}

\author{Jose Bonilla}
\affiliation{Department of Physics and Astronomy,
University of Tennessee, Knoxville, TN 37996, USA}

\author{Thomas R.~Richardson\,\orcidlink{0000-0001-6314-7518}}
\affiliation{Department of Physics, University of California, Berkeley, CA 94720, USA}
\affiliation{Nuclear Science Division, Lawrence Berkeley National Laboratory, Berkeley, CA 94720, USA}

\author{Sonia Bacca\,\orcidlink{0000-0002-9189-9458}}
\affiliation{Institut f\"ur Kernphysik and PRISMA$^+$ Cluster of Excellence, Johannes Gutenberg-Universit\"at, 55128 Mainz, Germany}
\affiliation{Helmholtz-Institut Mainz, Johannes Gutenberg Universit\"at Mainz, D-55099 Mainz, Germany}

\author{Chen Ji\,\orcidlink{0000-0002-4849-480X}}
\email{jichen@ccnu.edu.cn}
\affiliation{Key Laboratory of Quark and Lepton Physics, Institute of Particle Physics, Central China Normal University, Wuhan 430079, China}
\affiliation{Southern Center for Nuclear-Science Theory (SCNT),
Institute of Modern Physics, Chinese Academy of Sciences, Huizhou 516000, Guangdong Province, China}

\author{Lucas Platter\,\orcidlink{0000-0001-6632-8250}}
\affiliation{Department of Physics and Astronomy,
University of Tennessee, Knoxville, TN 37996, USA}
\affiliation{Physics Division, Oak Ridge National Laboratory, Oak Ridge, TN 37831, USA}

\begin{abstract}
 We calculate the nuclear-structure correction to the hyperfine splitting in both electronic and muonic deuterium using interactions from chiral effective field theory. We explore the sensitivity  to different parameterizations of the nucleon-nucleon force, 
 study the convergence pattern in the order-by-order chiral expansion, and estimate remaining uncertainties.
 Our results are consistent with earlier calculations from pionless effective field theory,  offering new insights for a robust uncertainty quantification. Thanks to the order-of-magnitude reduction in uncertainty achieved with chiral effective field theory, the two-photon exchange contribution in electronic deuterium agrees with experimental extractions within $0.7\sigma$, in contrast to the $2.7\sigma$ discrepancy observed in muonic deuterium.  This study  lays the groundwork for extending TPE calculations to HFS in heavier atomic systems.
\end{abstract}

\maketitle

\section{Introduction}
    \label{sec:introduction}
The hyperfine splittings (HFS) of atoms originate from the magnetic interactions between electrons (or other orbiting leptons) and the nucleus. High-precision measurements of HFS have become instrumental probes for nuclear-structure studies, offering detailed insights into the underlying nuclear interactions. In particular, studying HFS in electronic and muonic deuterium atoms provides an accurate way to investigate the  strong interactions that govern the nuclear structure at the MeV energy scale. This is achieved using high-precision laser spectroscopy at energy scales that are several orders of magnitude smaller than those of the nucleus.  Such studies have been successfully carried out for light electronic atoms \cite{hellwigMeasurementUnperturbedHydrogen1970, winelandAtomicDeuteriumMaser1972, rosnerHyperfineStructureState1970, guanProbingAtomicNuclear2020,Schneider:2022mze,Dickopf_2024} and muonic atoms \cite{antogniniProtonStructureMeasurement2013a, pohlLaserSpectroscopyMuonic2016a, thecremacollaborationHelionChargeRadius2023a}. When combined with accurate bound-state quantum electrodynamics (QED) calculations \cite{winelandAtomicDeuteriumMaser1972, i.eidesTheoryLightHydrogenlike2001}, these measurements enable stringent  tests of both nuclear-structure theories and QED theories.

Despite these advancements, accurate theoretical predictions for HFS in both electronic and muonic atoms are constrained by nuclear-structure effects, particularly those arising from two-photon exchange (TPE) processes. 
TPE contributions can be separated into elastic and inelastic components. 
The elastic TPE, characterized by the Zemach radius ($r_Z$), results from the convolution of the nuclear charge and magnetic distributions \cite{Zemach:1956zz, Friar:2003zg}. 
In contrast, the inelastic TPE, or nuclear polarizability, arises from excitations of the nucleus through electromagnetic probes \cite{friarNuclearCorrectionsHyperfine2005, friarNuclearPhysicsHyperfine2005}. 
These TPE effects are crucial in resolving discrepancies between experimental HFS measurements and QED-based theoretical predictions. 
For instance, in the 1S state of atomic deuterium (denoted with $e ^2$H) \cite{winelandAtomicDeuteriumMaser1972, i.eidesTheoryLightHydrogenlike2001}, the observed difference between experimental and QED-predicted HFS is
\begin{align}
\label{eq:tpe-D-exp}
\nu_{\rm exp}(e^2\text{H})-\nu_{\rm QED}(e^2\text{H}) = 45.2\;\text{kHz}.
\end{align}
For the 2S state of muonic deuterium (denoted with $\mu^2$H) \cite{Krauth:2015nja, kalinowskiNuclearstructureCorrectionsHyperfine2018} that difference is 
\begin{align}
\label{eq:tpe-muD-exp}
\nu_{\rm exp}(\mu^2\text{H}) - \nu_{\rm QED}(\mu^2\text{H}) = 0.0966(73)\;\text{meV}.
\end{align}
These discrepancies are dominated by the TPE contributions.  
These effects were traditionally calculated with the Low-term formalism~\cite{friarNuclearCorrectionsHyperfine2005, friarNuclearPhysicsHyperfine2005}, where the photon energy is set to zero in the TPE process. Using the closure relation, the TPE contribution is approximately calculated as a nuclear ground-state observable. 
The closure approximation was further extended in~\cite{kalinowskiNuclearstructureCorrectionsHyperfine2018} by partially including photon-energy-dependent contributions in TPE using perturbation theory. 
However, neither the original nor the extended Low-term formalism captures the correct energy-dependent weight in the TPE calculation, leading to an incomplete description.
Furthermore, it has recently been demonstrated that the closure approximation does not satisfy the requisite Ward identities \cite{Plestid:2025ojt}, thus yielding a model dependence in the calculated TPE effects.

To address these challenges, recent theoretical advancements have introduced an enhanced formalism for assessing  TPE effects on the HFS, meticulously integrating nuclear excitations and recoil effects \cite{jiNuclearStructureEffects2024}. 
The TPE contributions to the HFS in $^2$H and $\mu^2$H was studied by adopting nucleon-nucleon interactions from pionless effective field theory (\nopieft)~\cite{Hammer:2019poc}. 
This framework is particularly adept for systems where the typical momentum scale in the physical process is substantially lower than the pion mass. 
\nopieft offers a systematic low-energy expansion of the nuclear Hamiltonian, utilizing only contact interactions between nucleons with short-range physics at and beyond the scale of pion exchange embedded in a series of low-energy constants of contact interactions.
Its expansion parameter is the ratio of the interaction range ($R$) to the nucleon-nucleon scattering length ($a$), denoted as $R/a$. In Ref. \cite{jiNuclearStructureEffects2024}, TPE effects in $^2$H and $\mu^2$H were calculated at next-to-next-to-leading order  in this expansion, with the uncertainty due to EFT truncation estimated to be $\sim 5-6$~\%.

Nucleon-nucleon ($NN$) interactions have been derived from chiral effective field theory ($\chi$EFT) as well \cite{Weinberg:1990rz, Weinberg:1991um, Epelbaum:2012vx, Machleidt:2011zz, Hammer:2019poc}. This framework incorporates both, pion exchanges and contact interactions, into the nuclear forces, is  applicable  to a wider energy regime, and is widely used in many-body systems~\cite{baccapastore,Heiko}.
The expansion parameter in $\chi$EFT is $m_\pi/\Lambda_b$, where $m_\pi$ denotes the pion mass and $\Lambda_b$ denotes the chiral breakdown scale which has been inferred to be approximately 600~MeV \cite{Epelbaum:2014efa,Furnstahl:2015rha}. 
This higher breakdown scale compared to that in \nopieft ($\sim 1.4 m_\pi$~\cite{Ekstrom:2024dqr}) makes $\chi$EFT particularly well-suited for studying nuclear structure in heavier nuclei, where dynamics of pion exchanges play a significant role. 

In this study, we extend the formalism of Ref.~\cite{jiNuclearStructureEffects2024}, and employ $\chi$EFT to derive the necessary nuclear matrix elements for calculating  TPE contributions to the HFS in $^2$H and $\mu^2$H. 
By  including operators from chiral $NN$ interactions at various orders in the chiral expansion, and one-body electromagnetic currents, this work aims to determine  TPE contributions to the HFS with high accuracy and to benchmark results from $\chi$EFT against those from \nopieft. 
The comparative analysis not only validates the consistency of different EFTs, but also establishes a foundation for extending the calculations of  TPE effects in HFS to atomic systems involving heavier nuclei. 

The paper is structured as follows. In Sec.~\ref{sec:formalism}, we review the formalism derived in Ref.~\cite{jiNuclearStructureEffects2024} that is employed to extract the elastic and inelastic nuclear TPE contribution, and single-nucleon TPE contribution to the HFS. In Sec.~\ref{sec:results}, we present our numerical results and uncertainty estimation. Conclusion and discussion are given in Sec.~\ref{sec:conclusion}.

\section{Formalism}
    \label{sec:formalism}

The dominant HFS effects of the atomic $n s_{1/2}$ state, where $n$ is the principal quantum number, in $^2$H and $\mu^2$H are generated through the contact interaction between the magnetic moments of the lepton and the nucleus~\cite{Schwartz:1955pr,Woodgate:1983de,i.eidesTheoryLightHydrogenlike2001}
    \begin{equation}
            \label{eq:fermi_hamiltonian}
    	H_F = \frac{2\pi\alpha_{em} g_m}{3m_l M_N}\phi^{2}_{n}(0)\bm{\sigma}^{(l)}\cdot\bm{I}~ ,
    \end{equation}
where $m_l$ and $M_N$ are the lepton and nucleon masses, respectively, $g_m$ is the deuteron magnetic $g$-factor, and $\alpha_{em}$ is the electromagnetic fine structure constant. The vector Pauli matrix ${\bm \sigma}^{(l)}$ acts on the leptonic Hilbert space while $\bm I$ denotes the  total angular momentum operator (also called nuclear spin) in the nuclear part of the Hilbert space. The squared atomic wave function at the origin is given by
\begin{equation}
            \label{eq:atomic_wf}
        \phi^{2}_{n}(0) = \left(\frac{\alpha_{em} m_l m_d}{m_l+m_d}\right)^3 \frac{1}{n^3 \pi}~,
\end{equation}
where $m_d$ is the mass of the deuteron.
The HFS energy can be obtained at $O(\alpha_{em}^4)$ by taking the expectation value of the operator $H_F$ on the specific hyperfine state
    \begin{align}
            \label{eq:fermi_energy}
    	E_F &=  \bra{(ns_{1/2};\psi_d I)F,M_F}H_F\ket{(ns_{1/2};\psi_d I)F,M_F}
     \nonumber\\
     &= \frac{2\pi\alpha_{em} g_m}{3m_l M_N}\phi^{2}_{n}(0)
     \left[F(F+1)-I(I+1)-\frac{3}{4}\right],
    \end{align}
where $E_F$ is known as the Fermi-contact hyperfine energy, $F$, and $M_F$ are the total angular momentum and its $z$-projection of the atomic hyperfine structure state, respectively, and $\ket{\psi_d I}$ is the nuclear ground state with total angular momentum $I$; for the deuteron, $I=1$, and $F=1/2,3/2$.

\begin{figure}[hb]
    \centering    \includegraphics[width=0.5\textwidth]{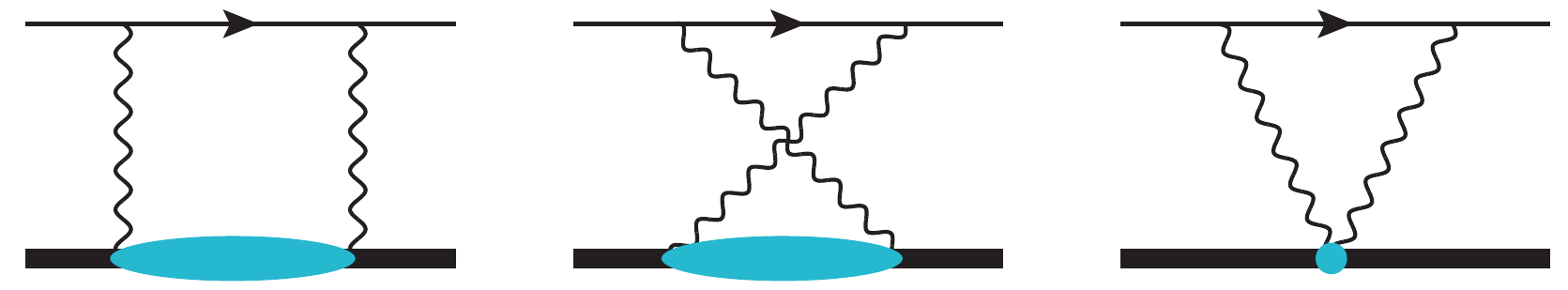}
    \caption{The lepton-nucleus two-photon-exchange box, cross and seagull diagrams.}
    \label{fig:tpe}
\end{figure}

At $O(\alpha_{em}^5)$, the hyperfine structure receives corrections from TPE between the nucleus and the bound lepton.
The operator that describes this process is given by the Feynman diagrams in Fig.~\ref{fig:tpe}, whose expression in Feynman gauge is given by
\begin{equation}
\label{eq:H2g}
	H_{2\gamma} \!\!=\!\! i(4\pi\alpha_{em})^2\phi^{2}_{n}(0)\!\!\int \!\!\frac{d^4q}{(2\pi)^4}\frac{\eta_{\mu\nu}(q)T^{\mu\nu}(q)}{(q^2+i\epsilon)^2(q^2-2m_lq_0+i\epsilon)}~,
\end{equation}
where $T^{\mu \nu}$ is the nuclear Compton tensor, known as the forward virtual Compton amplitude. Here, $\eta^{\mu \nu}$ is the leptonic tensor, whose full expression is given by
\begin{align}
\label{eq:eta-tensor}
\eta^{\mu \nu} =& (k-q)^\mu \delta^{\nu 0} +(k-q)^{\nu} \delta^{\mu 0} + q_0 g^{\mu \nu} 
\nonumber\\
&+ i q_0 \epsilon^{0 \mu \nu i}\sigma_i^{(l)} + i \epsilon^{\mu \nu i j} \sigma^{(l)}_i q_j~.
\end{align}
The hyperfine structure depends on the lepton spin and is therefore contained in the lepton-spin dependent part of $\eta^{\mu \nu}$, which is given by
    \begin{equation}
    \label{eq:tensor-l-reduce}
        \eta^{\mu \nu} \supset i q_0 \epsilon^{0 \mu \nu i} \sigma_i^{(l)} + i \epsilon^{\mu \nu ij}  \sigma_i^{(l)} q_j \, .
    \end{equation}
By inserting the expression above into Eq.~\eqref{eq:H2g}, we decompose the TPE operator into two contributions:
\begin{eqnarray}
\mathcal{H}_{\rm 2\gamma} &=& \mathcal{H}_{\rm 2\gamma}^{(0)} + \mathcal{H}_{\rm 2\gamma}^{(1)}~,\
\end{eqnarray}
\begin{eqnarray}
\label{eq:tim-cur}
\mathcal{H}_{\rm 2\gamma}^{(0)} &=& (4\pi\alpha_{em})^2 \phi^{2}_{n}(0)\\
\nonumber
&&\int \frac{d^4 q}{(2\pi)^4}
\frac{(\vec{\sigma}^{(l)} \times \q )_m  \left[ T^{m0}(q,-q) - T^{0m}(q,-q) \right] }{(q^2+i\epsilon)^2(q^2-2m_{l} q_0+i\epsilon )}~,
\end{eqnarray}
\begin{eqnarray}
\label{eq:cur-cur}
\mathcal{H}_{\rm 2\gamma}^{(1)} &=&- (4\pi\alpha_{em})^2 \phi^{2}_{n}(0)\\
\nonumber
&&\int \frac{d^4 q}{(2\pi)^4}
\frac{q_0\, \epsilon^{0 i j k}\sigma_k^{(l)} T^{ij}(q,-q)  }{(q^2+i\epsilon)^2(q^2-2m_{l} q_0+i\epsilon )}~,
\end{eqnarray}
where $T^{m0}$ and $T^{ij}$ are, respectively, the time-space and space-space components of the nuclear Compton tensor.
The elastic and inelastic contributions  are obtained by inserting a complete set of intermediate states in between the current operators contained in the Compton tensor. The projection on the ground state leads to the elastic term, while the projection on all excited states generates the inelastic (polarizability) contribution.

\begin{widetext}

The nuclear polarizability tensor operator from the combination of box and cross TPE Feynman diagrams can be written as
\begin{equation}
\label{eq:T-nucl-pol}
T_{\rm pol}^{\mu \nu}=
\SumInt_{\omega_{\boldsymbol q}} \left[\frac{ J^\mu (-\q)|\omega_{\q}\rangle \langle \omega_{\q}|  J^\nu  (\q) }{q_0-\omega_{\q} +i\epsilon} + \frac{ J^\nu (\q) |\omega_{-\q}\rangle \langle \omega_{-\q}| J^\mu(-\q) }{-q_0-\omega_{\q} +i\epsilon} \right]~,
\end{equation}
where $|\omega_{\q}\rangle$ indicates the nuclear intermediate states with total center-of-mass momentum $\q$ and excitation energy $\omega=E-E_0$, $\omega_{\q}=\omega+\q^2/(4M_N)$ denotes the energy difference between the ground state $\ket{\psi_d}$ in rest frame and the intermediate state $\ket{\omega_\q}$. 
The symbol $\SumInt_{\omega_{\boldsymbol q}}$ involves a sum over intermediate excited states of the nucleus, which we only consider to be only of \NN nature. 

Combining these ingredients and integrating out $q_0$ in Eqs.~(\ref{eq:tim-cur}) and (\ref{eq:cur-cur}), the nuclear polarizability operator is decomposed into two contributions,
\begin{align}
	\label{eq:H0pol}
	H^{(0)}_{pol} &= \frac{i\alpha_{em}^2 \phi^{2}_{n}(0)}{2\pi m^{2}_{l}}\int \frac{d^3q}{\q^4}h^{(0)}(\omega_{\q},q)\bm{\sigma}^{(l)}\cdot \{\q\times {\bf J}(-\q), J_0(\q)\}~,\\
	\label{eq:H1pol}
	H^{(1)}_{pol} &= \frac{i\alpha_{em}^2 \phi^{2}_{n}(0)}{2\pi m^{2}_{l}} \int \frac{d^3q}{\q^2}h^{(1)}(\omega_{\q},q)\bm{\sigma}^{(l)}\cdot [{\bf J}(-\q)\times {\bf J}(\q)]~,
\end{align}
where $J_0$ is the charge, 
and where from now on $q$ will be used to denote the magnitude of the three-momentum, i.e. $ q = \abs{\vb q}$.
The current operator can also be decomposed into convection and magnetic contributions
${\bf J}$=${\bf J_c}$+${\bf J_m}$. 
It should be noted that the anticommutator $\{\cdots\}$ in Eq.~\eqref{eq:H0pol} and the commutator $[\cdots]$ in Eq. \eqref{eq:H1pol} involve a sum over intermediate excited states between the two nuclear electromagnetic charge or current operators.
The kernels $h^{(0)}$ and $h^{(1)}$ are given by 
    \begin{align}
    \label{eq:h0}
	   h^{(0)}(\omega,q) & = \left( 2 + \frac{\omega}{E_q} \right) \left[ \frac{E^2_q + m^2_l + E_q\omega}{(E_q + \omega)^2 - m^2_l} \right] - \frac{2q + \omega}{q + \omega}~ \, , \\
       \label{eq:h1}
	   h^{(1)}(\omega,q) & = \frac{1}{E_q}\left[ \frac{E^2_q + m^2_l + E_q\omega}{(E_q + \omega)^2 - m^2_l} \right] - \frac{1}{q + \omega}~ \, ,
    \end{align}
and $E_q = \sqrt{\q^2 + m_l^2}$.

The nuclear polarizability corrections to the HFS are evaluated through the expectation value of the polarizability operators $H^{(0,1)}_{pol}$ for the atomic $ns$ states as
    \begin{equation}
            \label{eqref:E_01}
        E_\text{pol}^{(0,1)} = \bra{(ns_{1/2},\psi_d I)F,M_F}H^{(0,1)}_{pol}\ket{(ns_{1/2},\psi_d I)F,M_F} \, .
    \end{equation}
Similar to $H_F$ in Eq.\eqref{eq:fermi_hamiltonian},  $H^{(0,1)}_{pol}$ in Eqs.~\eqref{eq:H0pol} and \eqref{eq:H1pol} are also expressed as the inner product between $\bm{\sigma}^{(l)}$ and a vector operator. Using the Wigner-Eckart theorem,  $E_\text{pol}^{(0,1)}$ must be proportional to the leading Fermi energy $E_F$, with the coefficients independent of quantum number $F$ of the atomic hyperfine state. Therefore, we have
\begin{align}
	\label{eq:delta_0}
	E^{(0)}_{\rm pol} & =\frac{6\alpha_{em} M_N}{\pi m_lg_m I} E_F\int^{\infty}_{\omega_{th}} d\omega' \int^{\infty}_{0} dq~h^{(0)}(\omega',q)S^{(0)}(\omega',q)~, \\	
	\label{eq:delta_1}
	E^{(1)}_{\rm pol} &=  -\frac{6\alpha_{em} M_N}{\pi m_lg_m I} E_F\int^{\infty}_{\omega_{th}} d\omega' \int^{\infty}_{0} dq~h^{(1)}(\omega',q)S^{(1)}(\omega',q)\,,
\end{align}
and $S^{(0,1)}(\omega',q)$ are nuclear response functions given by
\begin{align}
	\label{eq:S0}
	S^{(0)}(\omega',q) &= \int \frac{d\hat{q}}{4\pi  q^2}
    \SumInt_{\omega} {\rm Im}\big( \bra{\psi_dI,I} J_0(-\q) \ket{\omega_\q} \bra{\omega_\q} \big[\q \times J_M(\q)\big]_3 \ket{\psi_dI,I} \big) \delta(\omega'-\omega_{\q} )~,\\
	\label{eq:S1}
	S^{(1)}(\omega',q) &= \int \frac{d\hat{q}}{4\pi}
    \SumInt_{\omega} \epsilon^{3jk} {\rm Im}\big( \bra{\psi_d I,I} J_c^j(-\q) \ket{\omega_\q} \bra{\omega_\q} J_m^k(\q) \ket{\psi_dI,I} \big) \delta(\omega'-\omega_{\q})~,
\end{align}
where $\ket{\psi_dI,I}$ denotes the deuteron ground state with total angular momentum $I$ and with maximal $z$-projection, i.e., $M_d=I$. 
The deuteron bound-state and continuum states can be rigorously solved by Lippmann-Schwinger equations represented in momentum space. 
Details can be found in Ref.~\cite{JoseThesis}.

For the elastic contribution, the nuclear tensor operator for the combination of box and cross TPE diagrams is written as
\begin{equation}
\label{eq:T-nucl-el}
T_{\rm el}^{\mu \nu} = \left[\frac{ J^\mu (-\q)|\psi_{d,\q}\rangle \langle \psi_{d,\q}|  J^\nu  (\q) }{q_0-\q^2/(4M_N) +i\epsilon} + \frac{ J^\nu (\q) |\psi_{d,-\q}\rangle \langle \psi_{d,-\q}| J^\mu(-\q) }{-q_0-\q^2/(4M_N) +i\epsilon} \right]~,
\end{equation}
where $|\psi_{d,\q}\rangle$ indicates the intemediate deuteron ground state boosted with a total center-of-mass momentum $\q$. 
Therefore, energy transfer for the elastic TPE process is carried purely by the nuclear recoil energy  $\q^2/(4M_N)$.
Similarly, inserting $T_{\rm el}^{\mu \nu}$ into Eqs.~\eqref{eq:tim-cur} and \eqref{eq:cur-cur} and using Wigner-Eckart theorem leads to the elastic TPE contribution 
\begin{align}
        E_{\rm el}^{(0)} & = \frac{2 \alpha_{em} E_F}{\pi m_l} \int dq \left[ h^{(0)}\left( \frac{q^2}{4 M_N}, q\right) F_d^{(C)}(q^2) F_d^{(M)}(q^2) - \frac{4 m_l m_r}{q^2} \right] \, , \label{eq:tpe_elastic0} \\
        E_{\rm el}^{(1)} & = - \frac{\alpha_{em} E_F}{2 \pi m_l M_N} \int dq \, q^2  h^{(1)}\left( \frac{q^2}{4 M_N}, q\right) F_d^{(M)}(q^2) F_d^{(M)}(q^2) \, , \label{eq:tpe_elastic1}
    \end{align}
where $F_d^{(C)}$ ($F_d^{(M)}$) is the charge (magnetic) form factor of the deuteron.
\end{widetext}

Both $E_{\rm el}^{(0)}$ and $E_{\rm pol}^{(0)}$ involve the charge-current correlated nuclear matrix elements, which give the leading contribution to nuclear TPE effects. In addition, $E_{\rm el}^{(1)}$ and $E_{\rm pol}^{(1)}$ involve the current-current correlated matrix elements, whose contributions are expected to be one-order higher in $(1/M_N)$ compared to the leading effects~\cite{friarNuclearCorrectionsHyperfine2005, friarNuclearPhysicsHyperfine2005}.  

Contributions from the seagull TPE diagrams are expected to be further suppressed and not included in this analysis. The charge-current seagull term is of relativistic order ($1/M^2$ higher). Although the current-current seagull term is non-relativistic, it is canceled in TPE due to crossing symmetry~\cite{friarNuclearCorrectionsHyperfine2005, friarNuclearPhysicsHyperfine2005}. 

Besides the TPE contribution due to the nuclear structure and dynamics, the complete TPE correction to hyperfine shift also receives single-nucleon contributions. When embedded in deuterium, the single-nucleon TPE is formulated into an expectation value of the deuteron ground state by~\cite{friarNuclearCorrectionsHyperfine2005,kalinowskiNuclearstructureCorrectionsHyperfine2018}
\begin{align}
E_{1N} =& -\frac{2\alpha_{em} m_l m_N E_F}{g_m(m_l+m_N)}   \langle \kappa_p  \tilde{r}_Z^p \vec{s}_p  + \kappa_n  \tilde{r}_Z^n  \vec{s}_n \rangle
\nonumber\\
=& -\frac{2\alpha_{em} m_l m_N E_F}{g_m(m_l+m_N)}   (\kappa_p  \tilde{r}_Z^p   + \kappa_n  \tilde{r}_Z^n  )(1-\frac{3}{2}P_D)
\label{eq:rzN},
\end{align}
where $\tilde{r}_Z^p$ and $\tilde{r}_Z^n$ represent the effective
proton and neutron Zemach radii, accounting for the full
single-nucleon TPE effects, including contributions from single-nucleon Zemach radii, recoil effects, and hadronic polarizabilities.
\cite{Tomalak:2019epja,Tomalak:2019prd,Antognini:2022arn,Hagelstein:2015lph,Hagelstein:2018bdi}.  $\kappa_p=2.793$ and $\kappa_n=-1.913$ are the magnetic moments of proton and neutron. $P_D$ denotes the $D$-wave probability of the deuteron ground state, which can be calculated in $\chi$EFT.
Therefore, the complete TPE contribution is
\begin{equation}
    E^{\rm HFS}_{\rm TPE} = E_{\rm el} + E_{\rm pol}+E_{1p} + E_{1n}~,
\end{equation}
where $E_\text{pol} =  E_{\text{pol}}^{(0)} + E_{\text{pol}}^{(1)} $ and $E_\text{el} =  E_{\text{el}}^{(0)} + E_{\text{el}}^{(1)} $.

\section{Results}
    \label{sec:results}

\begin{figure*}[t]
	\begin{center}
		\includegraphics[width = 0.45 \textwidth]{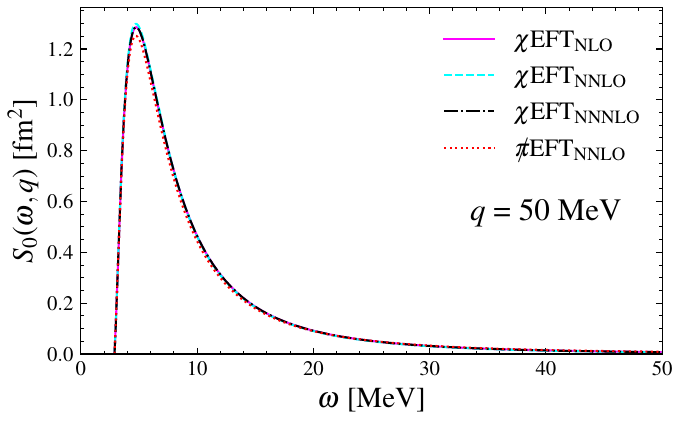}
		\includegraphics[width = 0.46 \textwidth]{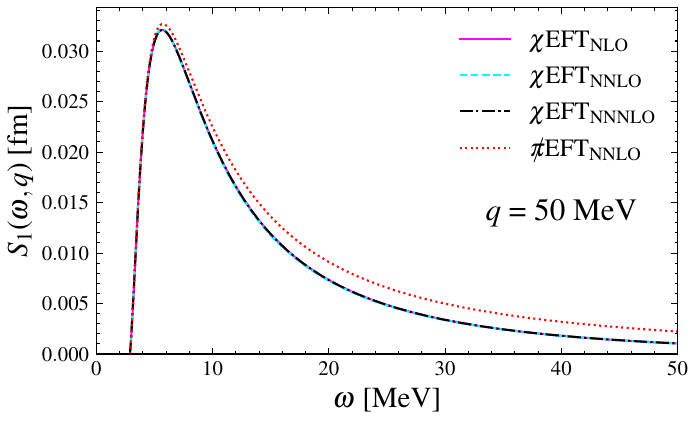}
		\includegraphics[width = 0.45 \textwidth]{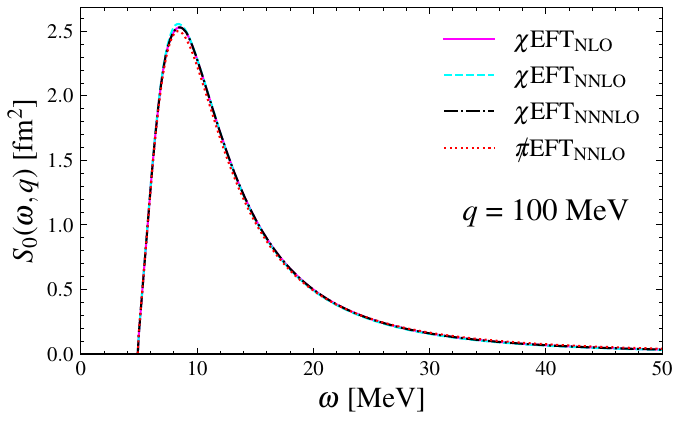}
		\includegraphics[width = 0.45\textwidth]{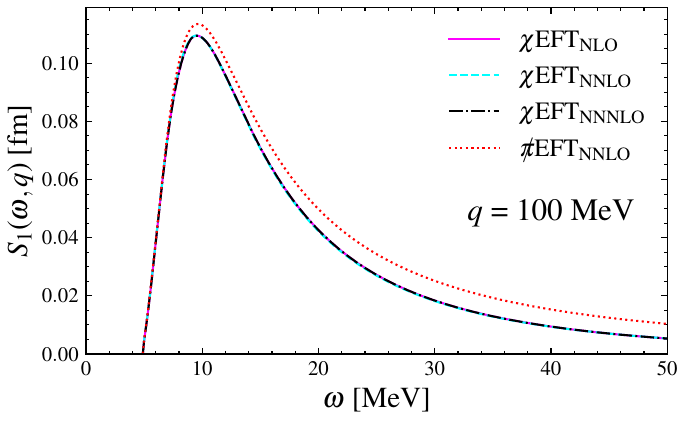}
		\caption{\label{fig:S1_upto_Fwave2} Response functions $S^{(0)}(\omega,q)$ (left column) and $S^{(1)}(\omega,q)$ (right column)  evaluated at two fixed momentum $q$ = 50 and 100 MeV plotted as a function of $\omega$. These results were obtained using $\chi$EFT interactions at \gls{nlo}, \gls{n2lo}, and \gls{n3lo} and with \nopieft interactions at \gls{n2lo}, with one-body  current operators and no nucleon form factors.        }
	\end{center}	
\end{figure*}

We start by analyzing  the polarizibility contributions to the HFS at different orders in the chiral EFT expansion.
In particular, we use the Bochum RS450 interactions of Ref.~\cite{Reinert:2017usi} truncated at \gls{nlo} and \gls{n2lo}, and 
the widely used Idaho interaction of Ref.~\cite{entemAccurateChargedependentNucleonnucleon2003} truncated at \gls{n3lo}. 
Note that Refs.~\cite{entemAccurateChargedependentNucleonnucleon2003,Reinert:2017usi} use different values for the pion-nucleon coupling constants: the RS450 potential uses the values obtained from a Roy-Steiner analysis~\cite{Hoferichter:2015hva,Siemens:2016jwj}, while the Idaho interaction uses coupling constants obtained from peripheral nucleon-nucleon scattering and pion-nucleon scattering. We will use these three parameterizations as the  \gls{nlo},  \gls{n2lo}, and  \gls{n3lo} chiral interactions in our subsequent  order-by-order truncation analysis.

Relevant matrix elements were calculated using only the one-body charge and vector current operator (see, e.g., Ref.~\cite{krebsNuclearCurrentsChiral2020a} for a detailed review). Two-body magnetic currents arising from one-pion exchange enter at \gls{nlo} in $\chi$EFT~\cite{Acharya:2020bxf}, and are of isovector nature. 
This operator connects the deuteron ground state with isospin 0 and spin 1, to a state with isospin 1 and spin 0. This transition is forbidden by the selection rules for the charge operator which cannot induce a spin change.
Therefore, for the deuteron the effect of the NLO two-body current cancels in $E_{\rm pol}^{(0)}$ for the current-charge interfering TPE process. A similar cancellation occurs in $E_{\rm pol}^{(1)}$.
Higher-order two-body charge and current operators that enter at \gls{n3lo} are omitted in this calculation.

In Fig.~\ref{fig:S1_upto_Fwave2}, we compare the response functions calculated with chiral interactions at three different orders, and those calculated with \nopieft at NNLO, for momentum transfer values of $q=$50 and 100 MeV. To highlight the differences due to the variation of the nucleon-nucleon force, the comparison is made in the point-nucleon limit.
We find that the response functions at \gls{nlo}, \gls{n2lo}, and \gls{n3lo} in the chiral expansion are very similar to each other and that additional orders do not seem to dramatically change the features of the response function already obtained at \gls{nlo}. This indicates that the chiral expansion is well converged. 
For the response functions calculated with \nopieft at NNLO, $S_0$ is consistent with the $\chi$EFT result, while $S_1$  displays small deviations from the $\chi$EFT result at large $\omega$ or $q$, which will be though suppressed by the integral kernels $h^{(0,1)}$ in Eqs.~\eqref{eq:delta_0} and \eqref{eq:delta_1}.
 For both response functions $S^{(0,1)}(\omega,q)$, 
 we have checked the contributions of individual partial waves and found that the $P$-wave  intermediate states dominate the strength.
  The contributions from other channels is very small at $q=50$ MeV, but not negligible in the total response function starting from $q=100$ MeV. We included partial waves up to angular momentum 3 ($F$-wave), which turns out to be enough to converge.

In Table~\ref{table:hfs_pol}, we show the results for $E_{\text{pol}}$ at the 
three different chiral orders with one-body electromagnetic charge and current operators.
Here, we include the nucleon form factors from Refs.~\cite{Lin:2021umk,Lin:2021umz,Lin:2021xrc}. 
The first two columns give the contribution for the 1S and 2S states of  $e^2$H.
The third and fourth column give the results for the 1S and 2S states in $\mu^2$H.

The systematic uncertainty from the $\chi$EFT expansion is estimated by incorporating the $Q$-expansion for $E_{\rm pol}$:
\begin{equation}
\label{eq:Qexp}
E_{\rm pol} = E_{\rm pol, NLO}^{(0)} \left(1+ \sum_{n=2}^{4} c_n Q^n \right),
\end{equation}
where $Q=m_\pi/\Lambda_b$ denotes the estimated chiral expansion parameter, with $m_\pi\approx 140\text{ MeV}$ being the pion mass scale, and $\Lambda_b\approx 600\text{ MeV}$ the $\chi$EFT breakdown scale.  According to the $\chi$EFT expansion, corrections at NLO, NNLO, and NNNLO are expected to be of orders $Q^2$, $Q^3$, and $Q^4$, respectively. 
Since $E_{\rm pol,NLO}^{(0)}$ already includes the LO contribution, it is the leading contribution in $E_{\rm pol}$ and goes as $Q^0$.   Contributions from $E_{\rm pol,NLO}^{(1)}$ are suppressed by $m_\pi/(2M_N)$ $E_{\rm pol}^{(0)}$, and are thus comparable to corrections of order $Q^2$. Based on this power counting, we can estimate the size of coefficients $c_i$ by
\begin{align}
c_2 Q^2 =& E_{\rm pol,NLO}^{(1)} / E_{\rm pol, NLO}^{(0)},\\
c_3 Q^3 = & E_{\rm pol,NNLO}^{(0)}/ E_{\rm pol, NLO}^{(0)}-1,\\
c_4 Q^4 = & (E_{\rm pol,NNNLO}^{(0)}+E_{\rm pol,NNLO}^{(1)})/ E_{\rm pol, NLO}^{(0)}
\nonumber\\
&-1-c_2Q^2-c_3 Q^3\,,
\end{align}
with coefficients $c_i$ expected of natural size. 
Using the values of $E_{\rm pol}$ calculated at each order, as summarized in Table~\ref{table:hfs_pol}, we determine the coefficients $c_i$ for the polarizabilities in both $e^2$H and $\mu^2$H, and list them in Table~\ref{table:ci}.
Since $E_{\rm pol}$ for 1S and 2S HFS states are only different by a scaling factor of 8, their $\chi$EFT truncation errors and coefficients $c_i$ are the same.
\begin{table}[ht]
	\setlength{\tabcolsep}{12pt}
    \renewcommand{\arraystretch}{1.2}
	\begin{tabular}{c c  c  c  }
    \hline
          &  c2 & c3 & c4 \\
          \hline
    $e^2$H &  0.989& 0.082 & 2.30\\
    $\mu^2$H & 0.610& 0.125 & 2.26\\
    \hline
    	\end{tabular}
    \caption{$\chi$EFT  expansion coefficients $c_i$ for analyzing truncation errors in $E_{\text{pol}}$.}
	\label{table:ci}
\end{table}

The uncertainty from the interaction truncation  at $\mathcal{O}(Q^5)$ is estimated as
 \begin{equation}
\Delta_{\text{pol}, NN}= E^{(0)}_\text{pol, NLO} Q^5\text{max} \{ 1, |c_2|, |c_3|,  |c_4|\}.
\label{deltaNN}
 \end{equation}
 This simple form has been proven to yield EFT truncation errors similarly to the Bayesian analysis approach~\cite{Epelbaum:2014efa,Furnstahl:2015rha,Acharya:2021lrv}. The uncertainty due to the neglected two-body currents (2BC)  at NNNLO is estimated to be of order $Q^4$, leading to 
 \begin{equation}
     \Delta_{\text{pol}, \rm{2BC}}= E^{(0)}_\text{pol, NLO} Q^4 \text{max} \{ 1, |c_2|, |c_3|\}.
     \label{delta2BC}
 \end{equation}
The $\chi$EFT truncation errors of the nuclear polarizability contributions are summarized in Table~\ref{table:hfs_pol}, with numbers in the first and second parentheses denoting respectively the uncertainties from higher-order $NN$ interactions and 2BC, namely from Eq.~(\ref{deltaNN}) and  Eq.~(\ref{delta2BC}), respectively.

\begin{table*}[ht]
	\setlength{\tabcolsep}{12pt}
    \renewcommand{\arraystretch}{1.2}
	\begin{tabular}{ l l  c  c  c  c}
		\Xhline{0.1em}
		& & $e{}^2$H (1S) & $e{}^2$H (2S) & $\mu {}^2$H (1S) & $\mu {}^2$H (2S) \\
        &
		& [kHz] &  [kHz] & [meV] & [meV] \\
        \Xhline{0.08em}
		\multirow{3}{*}{$E_{\text{pol}}^{(0)}$} & \gls{nlo} & 115.72 & 14.465 & 2.916 & 0.3645 \\
		& \gls{n2lo} & 115.60 & 14.450 & 2.911 & 0.3639 \\
		& \gls{n3lo} & 116.32 & 14.540 & 2.930 & 0.3662 \\
		\Xhline{0.08em}
		\multirow{3}{*}{$E_{\text{pol}}^{(1)}$} & \gls{nlo} & -6.23 & -0.779 & -0.097 & -0.0121 \\
		& \gls{n2lo} & -6.16 & -0.770 & -0.096 & -0.0120 \\
		& \gls{n3lo} & -6.15 & -0.769 & -0.096 & -0.0120 \\
		\Xhline{0.08em}
		\multirow{3}{*}{$E_{\text{pol}}$} & \gls{nlo} & 109.49 & 13.686 & 2.819 & 0.3524 \\
		& \gls{n2lo}& 109.44 & 13.680 & 2.816 & 0.3520 \\
		& \gls{n3lo} & 110.16 & 13.770 & 2.834 & 0.3543 \\
        \Xhline{0.08em}
        \multirow{2}{*}{$E_{\text{pol}}$} & \pieft & 110.16(17)(34) & 13.770(22)(43) & 2.834(4)(9) & 0.3543(6)(11) \\
          & \nopieft~\cite{jiNuclearStructureEffects2024} & 109.8(4.5) & 13.73(56) & 2.86(12) & 0.358(14) \\
		\Xhline{0.1em}
	\end{tabular}
    \caption{Deuteron $\chi$EFT  nuclear polarizablity contributions to HFS in $E_{\text{pol}}^{(0)}$ and $E_{\text{pol}}^{(1)}$, and their combined effects $E_{\text{pol}}$, at various chiral orders, compared to \pieft results (see text for details). 
}
	\label{table:hfs_pol}
\end{table*}

\begin{figure}
    \centering
    \includegraphics[width=0.45\textwidth]{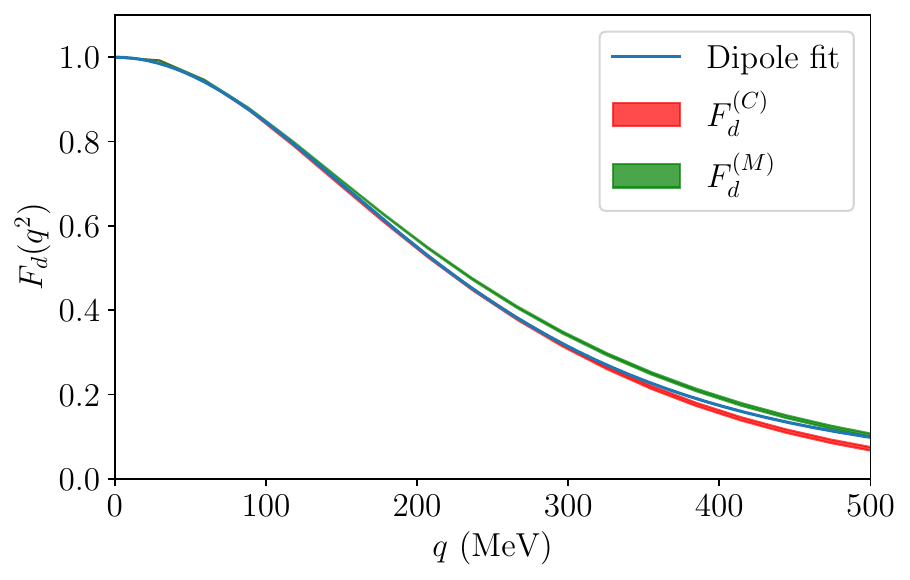}
    \caption{The red (green) band shows the charge (magnetic) form factor for the deuteron obtained in Ref.~\cite{Piarulli:2012bn}. The blue line shows a dipole fit of the form Eq.~\eqref{eq:dipole_fit} to the \gls{ChiEFT} result.}
    \label{fig:form_factor_fit}
\end{figure}

To obtain the elastic contributions, we make use of the deuteron electric and magnetic form factors from Refs.~\cite{Piarulli:2012bn}  using $\chi$EFT $NN$ interactions and currents at \gls{n3lo}.
To calculate the contributions in Eqs.~\eqref{eq:tpe_elastic0} and \eqref{eq:tpe_elastic1} to the HFS, it is convenient to fit the results of Ref.~\cite{Piarulli:2012bn} to a dipole parameterization 
	\begin{align}
		F^{(\rm dipole)}_d(\vb q^2) & = \frac{1}{(1 + a^2 \vb q^2)^2} \label{eq:dipole_fit}~.
	\end{align}
There is not a statistically significant difference in the fits of the electric and magnetic deuteron form factors; therefore, we adopt the same parameterization for both electric and magnetic deuteron form factors with $a = 0.002952(11) \ \text{MeV}^{-1}$.
The fit is shown with the $\chi$EFT calculations of Ref.~\cite{Piarulli:2012bn} in Fig.~\ref{fig:form_factor_fit}.

The parameterization in Eq.~\eqref{eq:dipole_fit} is sufficient to perform the integration in Eq.~\eqref{eq:tpe_elastic1}.
Once the kernels $h^{(0)}$ and $h^{(1)}$ are included in the integrand, the difference between the fit value and $\chi$EFT input is negligible.
On the other hand, the integral for $E_\text{el}^{(0)}$ in Eq.~\eqref{eq:tpe_elastic0} is numerically unstable.
Thus, we adopt a procedure similar to that used for the Zemach moments in Ref.~\cite{NevoDinur:2018hdo}.
To evaluate Eq.~\eqref{eq:tpe_elastic0}, we can split the integral into two regions; one for $q < 500$ MeV and one for $q > 500$ MeV.
In the first region $q < 500$ MeV, we evaluate the integral using a Guassian quadrature.
In the second region $q > 500$ MeV, the contribution to the integral from the first term in Eq.~\eqref{eq:tpe_elastic0} proportional to the form factors is negligible.
Therefore, we approximate the integrand by keeping only the second term, i.e, $-4 m_l m_r/q^2$ above $q = 500$ MeV and integrate this analytically.
However, we have also confirmed that the results are not sensitive to this specific choice of 500 MeV.
The results for the elastic contributions to the hyperfine energy shifts arising from this protocol are shown in the first two rows of Table~\ref{table:hfs_summary}.

The nucleon TPE contributions in Eq.~\eqref{eq:rzN}, require the effective nucleon Zemach radii, $\tilde{r}_Z^{p,n}$.  The proton TPE effects in both $e$H and $\mu$H have been determined using constraints from HFS spectroscopy measurements~\cite{Antognini:2022arn,Antognini:2022plb}. The neutron TPE effects were calculated using dispersive relations~\cite{Tomalak:2019epja,Tomalak:2019prd} based on scattering data.
These nucleon TPE effects are converted, using a scaling method, to $\tilde{r}_Z^{p,n}$ for electronic and muonic atoms~\cite{kalinowskiNuclearstructureCorrectionsHyperfine2018}, yielding:
\begin{align}
\label {eq:rZnp}
\tilde{r}_Z^{p,e} =& 0.883(2) \text{ fm}, &\tilde{r}_Z^{p,\mu} =& 0.906(2) \text{ fm},
\nonumber \\
\tilde{r}_Z^{n,e} =& 0.347(38) \text{ fm}, &\tilde{r}_Z^{n,\mu} =& 0.102(39) \text{ fm}.
\end{align}
 Eq.~\eqref{eq:rzN} also relies on the deuteron $D$-state probability $P_D$, whose values at LO, NLO, NNLO in RS450 potential, and NNNLO in Idahol potential are respectively $2.77$, $3.59$, $4.63$~\cite{Reinert:2017usi}, and $4.51$~\cite{entemAccurateChargedependentNucleonnucleon2003} in percentage. Similarly to Eq.~\eqref{eq:Qexp}, the uncertainty of $P_D$ from the truncated $NN$ interactions at $\mathcal{O}(Q^5)$ is estimated to be $\pm 0.073$ in percentage.
The uncertainties of single nucleon TPE contributions to HFS in $e^2$H and $\mu^2$H are mainly due to errors in the effective nucleon Zemach radii in~\eqref{eq:rZnp}, as shown in the first bracket for $E_{1p}$ and $E_{1n}$ in Table~\ref{table:hfs_summary}. The $\chi$EFT truncation error to $E_{1p}$ and $E_{1n}$ takes a minor effect, and is shown by the second bracket in \ref{table:hfs_summary}.

\begin{table*}[!t]
	\setlength{\tabcolsep}{12pt}
    \renewcommand{\arraystretch}{1.2}
	\begin{tabular}{l  c  c  c  c}
		\Xhline{0.1em}
		& $e{}^2$H (1S) & $e{}^2$H (2S) & $\mu {}^2$H (1S) & $\mu {}^2$H (2S) \\
        & [kHz] & [kHz]  & meV & meV   \\
        \Xhline{0.1em}
		$E_\text{el}^{(0)}$ & -39.54(11) & -4.943(14) & -0.9540(35)  & -0.1193(4) \\
        $E_\text{el}^{(1)}$ & -1.964(1) & -0.2455(1) & -0.01127(2) & -0.001409(3)\\
		$E_\text{pol}$
         & 110.16(38) & 13.770(48) & 2.834(10) & 0.3543(13)\\ 
        $E_{1p}$  & -33.14(7)(4) & -4.142(8)(5) & -0.949(2)(1) & -0.1186(2)(1) \\
		$E_{1n}$  & 8.95(93)(1) & 1.12(12)(0) & 0.075(28)(0) & 0.0093(37)(0)\\
        $\Delta_{3\gamma}$ & $\pm$0.49 & $\pm$0.061& $\pm$0.048& $\pm$0.006 \\
		\Xhline{0.08em}
		$E_{\text{TPE}}^{\text{HFS}}$ & 44.5(1.1) & 5.56(14) & 0.995(57) & 0.1243(72)\\
		\Xhline{0.08em}
		\nopieft~\cite{jiNuclearStructureEffects2024} & 41.7(2.6) & - & 0.940(73) & 0.118(9) \\
		Ref.~\cite{KHRIPLOVICH199613, Khriplovich2004} & 43 & - & - & - \\
		Ref.~\cite{friarNuclearCorrectionsHyperfine2005, friarNuclearPhysicsHyperfine2005}$_{\rm mod}$  & 64.5 & - & - & - \\
		Ref.~\cite{kalinowskiNuclearstructureCorrectionsHyperfine2018} & - & - & 0.304(68) & 0.0383(86) \\
		$\nu_{\rm exp}-\nu_{\rm QED}$~\cite{i.eidesTheoryLightHydrogenlike2001,Krauth:2015nja} & 45 & - & - & 0.0966(73) \\
		\Xhline{0.1em}
	\end{tabular}
    \caption{TPE contributions to HFS, decomposed into nuclear polarizability $E_{\rm pol}$, elastic $E_{\rm el}$, single proton  $E_{1p}$ and single neutron  $E_{1n}$ terms,
    compared with other theoretical and experimental results (see text for details). $\Delta_{3\gamma}$ denotes the estimated uncertainty from the missing 3PE effect. The original TPE results in Ref.~\cite{friarNuclearCorrectionsHyperfine2005, friarNuclearPhysicsHyperfine2005} are modified by adding the missing recoil and polarizability contributions to the single-nucleon TPE effects.}
	\label{table:hfs_summary}
\end{table*}

A summary of the obtained elastic and inelastic contributions to HFS  is shown in Table~\ref{table:hfs_summary}.
In the case of $E_{\rm pol}$, we now sum up the uncertainties from Eq.~(\ref{deltaNN}) and  Eq.~(\ref{delta2BC}) in quadrature.
We consider here also the single-nucleon contributions from the proton and the neutron. 
Our final results for the total TPE energy shift in the HFS are also reported in Table~\ref{table:hfs_summary}.

Relative to the TPE contribution, the effect of three-photon exchange (3PE) is expected to be suppressed by a factor of $\alpha_{em}$. For the 2S HFS in $\mu^2$H, the 3PE contribution was calculated to be $\approx -0.006$ meV~\cite{kalinowskiNuclearstructureCorrectionsHyperfine2018,Faustov:2014pra}. We include this value in the uncertainty budget to estimate the error due to the missing 3PE contributions. The 3PE corrections to HFS in $e^2$H have not been calculated in the literature. Therefore, we estimate that its contribution leads to an additional uncertainty of
\begin{equation}
\Delta_{3\gamma}[^2\text{H}(1S)] \approx \pm |\alpha_{em} E_{\rm TPE}| = \pm 0.49\;  \text{kHz}.	
\end{equation}
The total TPE contribution in this work is benchmarked with other theoretical results~\cite{jiNuclearStructureEffects2024,KHRIPLOVICH199613, Khriplovich2004,friarNuclearCorrectionsHyperfine2005, friarNuclearPhysicsHyperfine2005,kalinowskiNuclearstructureCorrectionsHyperfine2018} and compared to the difference between the experiments and the QED predictions~\cite{i.eidesTheoryLightHydrogenlike2001,Krauth:2015nja}. 
The TPE contributions to HFS in both $e^2$H and $\mu^2$H atoms, calculated within $\chi$EFT  are in very good agreement with the results shown in Ref.~\cite{jiNuclearStructureEffects2024} that were obtained using the pionless EFT interactions at \gls{n2lo}. For $E_{\rm el}$,  differences between $\chi$EFT and \nopieft calculations are 1.4\% and 1.9\% in $e^2$H and $\mu^2$H,  respectively. Differences in $E_{\rm pol}$ are smaller, i.e. 0.3\% and 1\% in $e^2$H and $\mu^2$H, respectively. $E_{1N}$ from $\chi$EFT is 6.8\% smaller than that from \nopieft due to the additional contribution of $P_D$. In \nopieft, $P_D$ only enters at NNNNLO from two insertions of $S$-to-$D$ mixing operators, and is thus not included in Ref.~\cite{jiNuclearStructureEffects2024}.  

The $E_{\rm TPE}^{\rm HFS}$ obtained in this work is overall consistent with Ref.~\cite{jiNuclearStructureEffects2024} within $1\sigma$ of the systematic combined uncertainty. Within $\chi$EFT,  uncertainties in $E_{\rm TPE}^{\rm HFS}$  are reduced by one order of magnitude compared to using \nopieft. Therefore, the dominant sources of uncertainty become the single-neutron TPE contributions and the 3PE effects. Comparing $E_{\rm TPE}^{\rm HFS}$ with the difference between spectroscopy measurements and QED predictions, namely with  $\nu_{\rm exp}-\nu_{\rm QED}$ in Eqs.~\eqref{eq:tpe-D-exp} and \eqref{eq:tpe-muD-exp}, the deviations in $e^2$H and $\mu^2$H are  $0.7\sigma$ and $2.7\sigma$, respectively.

\section{Conclusion}
    \label{sec:conclusion}

In this work, we have calculated the nuclear TPE contributions to the HFS in electronic and muonic deuterium in a $\chi$EFT framework using $NN$ interactions up to \gls{n3lo} and electromagnetic one-body currents. We based our estimate of the nuclear-theory uncertainty due to the $\chi$EFT truncation on the omitted $NN$ interactions at NNNNLO and the omitted two-body electromagnetic currents at NNNLO.
Our results are consistent with those of in Ref.~\cite{jiNuclearStructureEffects2024}. The   elastic and inelastic nuclear TPE contributions in  $\chi$EFT  show remarkable agreement with the \nopieft results. This agreement is due to the shallow binding of the deuteron, making physics of momentum scale at and beyond pion exchanges contribute less to the TPE correction.
However, we expect notable deviations between calculations in $\chi$EFT and pionless EFT for TPE contributions to the HFS in other atomic systems with heavier nuclei, where the mechanism of pion exchanges become more important. For the deuteron, only a small deviation in the single-nucleon TPE is observed because the $D$-state probability is included at all orders in $\chi$EFT but truncated up to NNNLO in \nopieft.

Using $\chi$EFT, nuclear-theory uncertainties are reduced by one order of magnitude compared to \nopieft, so that the total uncertainty is now dominated by the single-neutron TPE effects and by the omitted three-photon-exchange effects.

The predicted TPE contribution to $e^2$H HFS agrees with that extracted from the spectroscopy-QED deviation within $0.7\sigma$. Atomic HFS spectroscopy can therefore clearly serve as a sensitive probe of the nuclear magnetic structure of a nucleus. However, the $2.7\sigma$ discrepancy between the predicted and experimental TPE contribution in $\mu^2$H HFS requires further investigation of higher-order nuclear and QED effects.
It is also imperative to extend this calculation to heavier systems that are also accessible experimentally. Benchmarking theoretical predictions for TPE contributions to HFS against experiments in different atomic systems enables tests and constraints on parameters of nuclear interactions and electromagnetic currents.
Specifically, the ground-state HFS of $e^{3}$He has recently been measured with Penning-trap techniques~\cite{Schneider:2022mze}. A further measurement of HFS in $\mu^{3}$He is also planned~\cite{PSI_proposal}.

In addition, a better determination of the one-nucleon TPE contribution is desirable for improving the uncertainty of the toal TPE prediction. A persistent 
order-of-magnitude discrepancy exists between calculations of proton polarizability from chiral perturbation theory ($\chi$PT)~\cite{Antognini:2022arn,Hagelstein:2015lph,Hagelstein:2018bdi} and those from dispersion analysis~\cite{Tomalak:2019epja,Tomalak:2019prd}. Although this issue does not affect our input for the single-proton TPE contribution, as constrained by HFS spectroscopy, it suggests a potential underestimation  of the uncertainty in the single-neutron TPE effect, derived using dispersion analysis. Resolving this issue requires either higher-order $\chi$PT calculations of single-nucleon TPE effects, or an alternative approach using HFS in heavier systems, which would demand both more precise measurements and improved nuclear-structure calculations within $NN$ interactions at higher orders in $\chi$EFT.

\acknowledgments
We would like to thank Bijaya Acharya and Simone Li Muli for helpful discussions. 
T.~R.~R. thanks Andr\'e Walker-Loud for useful discussion.
This work was supported by the National Natural Science
Foundation of China (Grant Nos. 12175083, 12335002, and 11805078), the National Science Foundation (Grant Nos. PHY-2111426 and PHY-2412612), the Office of Nuclear Physics, the US Department of Energy (Contract No. DE-AC05-00OR22725), and  the Deutsche Forschungsgemeinschaft (DFG) through the Cluster of Excellence ``Precision Physics, Fundamental Interactions, and Structure of Matter'' (PRISMA${}^+$ EXC 2118/1) funded by the DFG within the German Excellence Strategy (Project ID 39083149).
T.~R.~R. is supported through the NSF through cooperative agreement 2020275.

%

\end{document}